\documentclass[aps,twocolumn,prd,nofootinbib]{revtex4}
\usepackage{amsmath}
\usepackage{graphicx}
\usepackage{dcolumn}
\usepackage{bm}
\usepackage{amssymb}
\usepackage{latexsym}

\begin{document}

\title{Probing Noncommutativity with Inflationary Gravitational Waves}

\author{Yi-Fu Cai$^a$\footnote{Email Address: caiyf@mail.ihep.ac.cn} and Yun-Song Piao$^{b}$\footnote{Email Address: yspiao@gucas.ac.cn}}
\affiliation{${}^a$Institute of High Energy Physics, Chinese
Academy of Sciences, P.O. Box 918-4, Beijing 100049, P. R. China}
\affiliation{${}^b$College of Physical Sciences, Graduate School
of Chinese Academy of Sciences, Beijing 100049, China}

\begin{abstract}

In this paper we study the behaviour of gravitational wave
background (GWB) generated during inflation in the noncommutative
field approach. From this approach we derive one additional term,
and then we find that the dispersion relation of the gravitational
wave would be modified and the primordial gravitational wave would
obtain an effective mass. Therefore it breaks local Lorentz
symmetry. Moreover, this additional term suppresses the energy
spectrum of the GWB greatly at low energy scale where the wave
length is near the current horizon. Due to this, a sharp peak is
formed in the energy spectrum at the low frequency band. This peak
should be a key criterion in detecting the spacetime
noncommutativity and a critical test for Lorentz symmetry breaking
in local field theory.

\end{abstract}

\maketitle

\hskip 1.6cm PACS number(s): {98.80.Cq, 04.30.-w} \vskip 0.4cm

\section{Introduction}

Under the standard model of cosmology plus the theory of
inflation\cite{Guth81,Steinhardt82,Linde82}, it is very natural to
predict the existence of the gravitational wave background. Scalar
type and tensor type primordial perturbations are generated during
inflation. The primordial scalar perturbations can be tested in
current observations of cosmic microwave
background\cite{CMBobserve,WMAP}. And they provide seeds for the
large scale structure(LSS), which then gradually forms today's
galaxies\cite{Mukhanov81,Guth82,Hawking82,Starobinsky82,Bardeen83}.
The tensor perturbations escape the horizon during inflation, and
then they keep conserved and form the relic GWB which carries the
information of the very early universe. There have been a number
of detectors operating for the signals of primordial GWB, e.g. Big
Bang Observer (BBO)\cite{BBO}, Planck\cite{CMBnextobserve}.
Besides, there have been evidences from the indirect detections of
the next generation of CMB observations, see related
analysis\cite{VPJ,Boyle06,Smith06}. The basic mechanism for the
generation of primordial GWB in cosmology has been discussed in
Refs. \cite{Grishchuk75,Allen88}. See \cite{Lyth93,Starobinski79}
for the gravitational waves generated during the epoch of
inflation; see \cite{Gasperini93} for that in Pre Big Bang
scenario; see \cite{Boyle04} for that in cyclic universe; see
\cite{piao06} for that from phantom super-inflation (also see
\cite{piao03,pin-others}).

Generally speaking, the GWB was generated in the epoch when the
energy scale of our universe is extremely high. Thus we should
take into consideration on more fundamental theories in logic,
namely, the string theory. In string theory it is commonly
realized that the spacetime coordinates of a Dp-brane is
noncommutative under certain external fields, and this can be
described by the language of noncommutative
geometry\cite{Douglas97,Seiberg99,Bigatti00,Alekseev00,Chu02}.
Accordingly, there are a lot of new properties, opening up new
challenges for current theories and experiments. For example,
causality is possibly violated as a consequence of the
noncommutative behaviour (see Refs. \cite{Seiberg00,Gomis00}), and
thus the Lorentz invariance\cite{Carroll01,Carson02,Chaichian04}
is also probably broken. When the noncommutativity is taken into
account during the very early universe, it would provide large
corrections to inflation, the primordial
fluctuations\cite{Brandenberger00,HuangLi,Cai04,HuangZhang},
black-body spectra of CMB\cite{Fatollahi}, and also primordial
magnetic field\cite{BY}. Furthermore, the paper \cite{Carmona02}
suggested another mechanism, named noncommutative field approach,
which can also lead to the Lorentz invariance breaking under
noncommutative fields while leaving spacetime commutative. The key
idea of this mechanism is to construct an effective action which
contains noncommutative terms partly obtained from noncommutative
geometry. Later, this noncommutative field approach has been
generalized to various models and has made many interesting
predictions (see Refs.
\cite{Gamboa05,Nascimento06,Arias06,Ferrari06}). If Lorentz
symmetry is indeed broken in our world, it would result in the
breaking of the CPT symmetry, which is possible to be detected in
experiments.
The paper \cite{Feng06} has already found some evidences to support
that cosmic CPT symmetry may not be conserved from the analysis of
data fitting for cosmological observations.

In our paper, using the noncommutative field approach, we
investigate the gravitational wave background during inflation.
The outline of this paper is as follows. In section II, we
introduce the additional term by applying the noncommutative field
approach to the equation of motion for the gravitational
perturbations in the flat Friedmann-Robertson-Walker(FRW)
background. Then we quantize the noncommutative tensor
perturbations and discuss the consistent initial conditions for
primordial gravitational waves with the noncommutative terms.
After that, we derive the classical solution of primordial tensor
perturbation with noncommutativity, and analyze the behaviour of
the primordial power spectrum and the corresponding spectral
index. In section III, from considering the influence of possible
factors on the energy spectrum, we discuss the transfer function
for the noncommutative GWB in order to relate the current GWB we
may probe with the primordial one. Finally, we give the
implications of the results from calculating the noncommutative
GWB in section IV, and compare them with the future development of
the cosmological observations. Then we summarize the fruitful
behaviors of GWB possessing noncommutativity which may be detected
in the future probe.

In this paper, we will see that, the dispersion relation of the
primordial gravitational wave will be modified, and its power
spectrum will exhibit difference from the normal one around the
CMB scale due to the noncommutative term. Furthermore, we can
obtain the same viewpoint from the relationship between the tensor
spectral index and its frequency. From the modified primordial
power spectrum, we can show that there will be a sharp peak in the
CMB scale of the energy spectrum observed today. In order to add
the contribution of the transfer function, we discuss some leading
corrections to the primordial GWB during the evolution of the
universe. Among these corrections, we mainly consider the
suppression of the GWB from the redshift, the impression of the
background equation of state of the universe when the GWB
re-enters the horizon, and the damping effects from the
anisotropic stress generated from some particles' freely
streaming\cite{Weinberg03,Pritchard04,Bashinsky,Dicus05}. After
considering all these contributions, we can clearly find that the
effects of noncommutative terms still play a significant role in
the low frequency range. So the results and predictions in the
above sections are important for future observations.

\section{Noncommutative tensor perturbations during inflation}

\subsection{Bases and Conventions}

In this section, we take a brief review of the standard theory of
tensor perturbation during inflation(for more details, see the
review article Ref. \cite{Brandenberger92}). In the flat FRW
background, the tensor perturbation is given in the metric,
\begin{eqnarray}
ds^2=a(\tau)^2[-d\tau^2+(\delta_{ij}+\bar h_{ij})dx^idx^j]~,
\end{eqnarray}
where $^{(0)}g_{\mu\nu}=diag(-a^2,a^2,a^2,a^2)$ is the background
metric, $\tau$ is the conformal time, $a(\tau)$ is the scale
factor, and the Latin indexes represent spatial coordinates. Here
the perturbation $\bar h_{ij}$ satisfies the following
constraints:
\begin{eqnarray}
\bar h_{ij}=\bar h_{ji}~;~\bar h_{ii}=0~;~\bar h_{ij,j}=0~.
\end{eqnarray}
Therefore, $\bar h_{ij}$ only have two degrees of freedom
corresponding to two polarizations of gravitational waves.

To start from Einstein's field equation, one can deduce the action
of free tensor perturbation as follows:
\begin{eqnarray}
^{(2)}{\cal S}_g=\int d\tau d^3x \frac{1}{64\pi G}a^2[\bar
h_{ij}'^2-(\partial_l{\bar h_{ij}})^2]~,
\end{eqnarray}
where prime represents the derivative with respect to the conformal
time. The interaction part of the action with other matter sources
is of the form
\begin{eqnarray}
^{(2)}{\cal S}_m=\int d\tau d^3x\frac{1}{2}a^4\sigma_{ij} \bar
h_{ij}~,
\end{eqnarray}
Here $\sigma_{ij}$ is the anisotropic part of the stress tensor,
constructed by the spatial components of the perturbed
energy-momentum tensor ${T^i}_j$,
\begin{eqnarray}
\sigma_{ij}=T^i_j-^{(0)}p\delta^i_j~.
\end{eqnarray}
Then one can derive the general equation of motion for tensor
perturbation:
\begin{eqnarray}
\bar h_{ij}''+2\frac{a'}{a}\bar h_{ij}'-{\bf \nabla}^{2}\bar
h_{ij}-16\pi G a^{2}\sigma_{ij}=0~.
\end{eqnarray}

The Fourier transformation of the tensor perturbation and the
anisotropic stress tensor takes the form,
\begin{eqnarray}
\bar h_{ij}(\tau, {\bf x})=\sqrt{16\pi
G}\int\frac{d^3k}{(2\pi)^\frac{3}{2}}H_{ij}(\tau, {\bf k})e^{i{\bf
k}{\bf x}}~,\\
\sigma_{ij}(\tau, {\bf x})=\sqrt{16\pi
G}\int\frac{d^3k}{(2\pi)^\frac{3}{2}}\Sigma_{ij}(\tau, {\bf
k})e^{i{\bf k}{\bf x}}~,
\end{eqnarray}
where we leave the Latin indices in order to keep the following
progresses to be general. In cosmology, what we care about is the
distribution of the spectra of gravitational waves and the
corresponding spectral index. Based on the above formalism, the
tensor power spectrum can be written as,
\begin{eqnarray}\label{tensor power}
P_T(k,\tau)&\equiv\frac{d\langle0|\bar{h}_{ij}^{2}|0\rangle}{d\,{\rm
ln}\,k}=32\pi G\frac{k^3}{(2\pi)^2}|H_{ij}(\tau, {\bf k})|^2~,
\end{eqnarray}
and the definition of tensor spectral index $n_T$ is given by
\begin{eqnarray}
n_T\equiv\frac{d\,{\rm ln}\,P_T}{d\,{\rm ln}\,k}~.
\end{eqnarray}

The GWB we observed today is characterized by the energy spectrum,
\begin{eqnarray}\label{tensor energy}
\Omega_{GW}(k,
\tau)\equiv\frac{1}{\rho_c(\tau)}\frac{d\langle0|\rho_{GW}(\tau)|0\rangle}{d\,{\rm
ln}\,k}~,
\end{eqnarray}
where $\rho_{GW}(\tau)$ indicates the energy density of
gravitational waves, and the parameter $\rho_c(\tau)$ is the
critical density of the universe. Make use of the Friedmann equation
\begin{eqnarray}
H^2(\tau)=\frac{8\pi G}{3}\rho_c(\tau)~,
\end{eqnarray}
the energy spectrum of GWB can be written as,
\begin{eqnarray}\label{approximate tensor energy0}
\Omega_{GW}(k,\tau)=\frac{8\pi
G}{3H^2(\tau)}\frac{k^3}{2(2\pi)^2}\frac{1}{a^2}\left(|H_{ij}'|^2+k^2|H_{ij}|^2\right).
\end{eqnarray}
In respect that the GWB we observed has already re-entered the
horizon, its mode should oscillate in the form of sinusoidal
function. Accordingly, we can deduce the relation between the power
spectrum and the energy spectrum we care about as follows
\begin{eqnarray}\label{approximate tensor energy}
\Omega_{GW}(k,\tau)\simeq\frac{1}{12}\frac{k^2}{a^2(\tau)H^2(\tau)}P_T(k,
\tau)~.
\end{eqnarray}
Note that, the equations (\ref{approximate tensor energy0}) and
(\ref{approximate tensor energy}) can be extended into the case of
noncommutative gravitational waves because in the framework of
noncommutative GWB the Hamiltonian is not modified, we will see the
detailed analysis in the next section.

\subsection{Quantization and Noncommutativity}

In this section, we will quantize the tensor perturbation, and
then introduce the noncommutative term in the canonical
commutation relation. Based on this, we will obtain a modified
theory of gravity in weak gravity approximation and finally
provide the equation of motion of primordial gravitational waves
in this theory. Since the action of tensor perturbation in
noncommutative field approach will break the Lorentz symmetry in
local Minkowski spacetime which can be exhibited in the dispersion
relation of gravitons\cite{Ferrari06}, we can see the
corresponding behaviour of gravitons in the the flat FRW universe
background.

Firstly, for simplicity, we redefine the tensor perturbation as
follows
\begin{eqnarray}\label{tensor redefine}
H_{ij}(\tau, {\bf k})\equiv\frac{h_{ij}(\tau, {\bf k})}{a(\tau)}~.
\end{eqnarray}
Therefore, in the momentum space the action of tensor perturbation
including the minimal coupling with matter is given by
\begin{eqnarray}
^{(2)}{\cal S}&=&\int
d\tau d^3k~[\frac{1}{4}(h_{ij}'^2+\frac{a''}{a}h_{ij}^2-k^2h_{ij}^2)\nonumber\\
&+&8\pi Ga^3\Sigma_{ij}h_{ij}]~,
\end{eqnarray}
with the equation of motion,
\begin{eqnarray}\label{old h equation}
h_{ij}''-\frac{a''}{a}h_{ij}+k^2h_{ij}-16\pi Ga^3\Sigma_{ij}=0~.
\end{eqnarray}
Secondly, from the standard canonical quantization, the conjugate
momentum can be written as,
\begin{eqnarray}
p_{ij}=\frac{\delta^{(2)}{\cal S}}{\delta
h_{ij}'}=\frac{1}{2}h_{ij}'~,
\end{eqnarray}
and the canonical Hamiltonian density in Fourier space is given by
\begin{eqnarray}
{\cal H}&=&p_{ij}h_{ij}'-{\cal L}\nonumber\\
&=&\frac{1}{4}p_{ij}^2-\frac{1}{4}\frac{a''}{a}h_{ij}^2+\frac{1}{4}k^2h_{ij}^2-8\pi
Ga^3\Sigma_{ij}h_{ij}~.
\end{eqnarray}
Converting the fields $h_{ij}$ and the conjugated momenta $p_{ij}$
into operators, we obtain the equal-time commutation relations from
the Poisson algebra,
\begin{eqnarray}\label{commutation relation1}
[h_{ij}(\tau, {\bf k}),h_{kl}(\tau, {\bf
k}')]&=&0~,\\\label{commutation relation2} [p_{ij}(\tau, {\bf
k}),p_{kl}(\tau, {\bf k}')]&=&0~,\\\label{commutation relation3}
[h_{ij}(\tau, {\bf k}),p_{kl}(\tau, {\bf
k}')]&=&\frac{1}{2}(\delta_{ik}\delta_{jl}+\delta_{il}\delta_{jk})\nonumber\\
&\times&\delta^{(3)}({\bf k}-{\bf k'})~.
\end{eqnarray}
Finally, following the similar formalism in \cite{Ferrari06}, we
make a deformation of the commutation relation of the conjugated
momenta into the noncommutative form
\begin{eqnarray}
[p_{ij}(\tau, {\bf k}),p_{kl}(\tau, {\bf
k}')]=\alpha_{ijkl}\delta^{(3)}({\bf k}-{\bf k'})~,
\end{eqnarray}
where $\alpha_{ijkl}$ is a constant matrix. In order to preserve
the properties of tensor perturbation, we require that
$\alpha_{ijkl}$ satisfies the following relation,
\begin{eqnarray}
\alpha_{ijkl}=\alpha_{jikl}=-\alpha_{klij}.
\end{eqnarray}
This is satisfied if
\begin{eqnarray}
\alpha_{ijkl}=\alpha_{ik}\delta_{jl}+\alpha_{il}\delta_{jk}+\alpha_{jk}\delta_{il}+\alpha_{jl}\delta_{ik},
\end{eqnarray}
with $\alpha_{jk}$ a constant antisymmetric matrix. One can check
that $\alpha_{ijkl}$ keeps the traceless property of tensor
perturbation. 
Moreover, in order to simplify the calculation,
$\alpha_{jk}$ can be further expressed as,
\begin{eqnarray}
\alpha_{jk}=-\epsilon_{0jkl}\alpha^l,
\end{eqnarray}
where $\epsilon_{ijkl}$ is a totally antisymmetric tensor, and we
use the convention that $\epsilon_{0123}=-1$ and
$\epsilon^{0123}=+1$ in our paper.

In order to recover a canonical system quantized in normal
commutative relations like eq. (\ref{commutation relation2}), we
construct a new conjugated momentum
$\pi_{ij}=p_{ij}-\frac{1}{2}\alpha_{ijkl}h_{kl}$ for the tensor
fields $h_{ij}$. Since in the derivation above we have discarded
the components of metric perturbations containing the time index,
the origin form of lagrangian is not covariant
. 
We modify the lagrangian by adding an additional term of covariant
form and then leave physical degrees of freedom in the following,
\begin{eqnarray}
\Delta{\cal
L}&=&-2\epsilon^{\rho\mu\nu\sigma}\alpha_{\sigma}h_{\nu\kappa}\partial_{\rho}h^{\kappa}_{\mu}~,\\
&\rightarrow&-2\alpha_0\epsilon^{0ijk}h_{kl}\partial_ih_{lj}-2\alpha_m\epsilon^{0jkm}h_{kl}h_{lj}'~,
\end{eqnarray}
where we call $\alpha_\mu$ as noncommutative parameters(NP).

Note that after making the Fourier transformation, the term
containing $\alpha_0$ disappears because of the symmetry of
exchanging indices ``$k$" and ``$j$". Here one can check that
Hamiltonian still keep invariant since the term containing
$\alpha_m$ can be cancelled by modified conjugated momentum and
the term including $\alpha_0$ also disappears due to the symmetry
of exchanging indices. Therefore, the modified Einstein-Hilbert
action of tensor perturbations is given by
\begin{widetext}
\begin{eqnarray}
^{(2)}{\cal S}^{new}=\int d\tau
d^3k~[\frac{1}{4}(h_{ij}'^2+\frac{a''}{a}h_{ij}^2-k^2h_{ij}^2)
-2\alpha_m\epsilon^{0jkm}h_{kl}h_{lj}' +8\pi
Ga^3\Sigma_{ij}h_{ij}]~,
\end{eqnarray}
\end{widetext}
and the corresponding equation of motion is of form
\begin{eqnarray}\label{h equation}
h_{ij}''+k^2h_{ij}-\frac{a''}{a}h_{ij}+8\alpha_m\epsilon^{0lim}h_{jl}'=16\pi
Ga^3\Sigma_{ij}~.
\end{eqnarray}
Compared with eq. (\ref{old h equation}), one can already find the
difference appeared in each one's dispersion relation roughly.

\subsection{Noncommutative Tensor Perturbations During Inflation}

In order to analyze the noncommutative tensor perturbations
explicitly, we need to obtain the classical solution of each mode of
tensor perturbations under the slow roll inflation background. Here
we choose NP of form: $\alpha_m=\{0,0,\alpha_3\}$ and substitute
this into Eq. (\ref{h equation}). Besides, we neglect the
anisotropic tensor fluctuations of matters and thus
$\Sigma_{ij}\simeq0$ here. According to the qualities of gauge
invariance, we can obtain only two independent modes which can be
expressed as
\begin{eqnarray}
v_1=\frac{1}{\sqrt{2}}(h_{11}+ih_{12})~,~~v_2=\frac{1}{\sqrt{2}}(h_{11}-ih_{12})~,
\end{eqnarray}
and in our note we call them left-handed and right-handed
respectively. These two modes satisfy their own equations:
\begin{eqnarray}\label{v1}
v_1''+8i\alpha_3v_1'+k^2 v_1-\frac{a''}{a}v_1=0~,\\\label{v2}
v_2''-8i\alpha_3v_2'+k^2 v_2-\frac{a''}{a}v_2=0~.
\end{eqnarray}

To solve these two equations explicitly, we need to know their
initial conditions. In the noncommutative field approach the
dispersion relation of gravitational waves has been modified, thus
it seems that the initial conditions should be different from
those in commutative case. Thus we need to reconsider the choice
of initial conditions. Here we fix the initial conditions by
requiring the initial state to be the lowest energy state(more
general initial conditions of modified dispersion relation, see
Ref. \cite{BrandenbergerM00}).
Since the noncommutative term does not contribute to energy
density, we can use the original form appeared in Eq.
(\ref{approximate tensor energy0}), and then give the expression
of the energy density:
\begin{eqnarray}
\rho_{GW}&=&\int_0^\infty
dk\frac{k^2}{8\pi^2a^4}\sum_{i=1,2}[v_i'{v_{i}^*}'-\frac{a'}{a}v_i'{v_{i}^*}-\frac{a'}{a}v_i{v_{i}^*}'\nonumber\\
&+&\frac{a'^2}{a^2}v_i{v_{i}^*}+k^2v_i{v_{i}^*}]~,
\end{eqnarray}
where we sum  'i' over the left-handed and right-handed modes.
Note that, the energy density can be divided into two part, for
one only containing $v_1$ and for the other only containing $v_2$.
Accordingly, we can analyze each mode separately and then obtain
the initial conditions for each one. Now we define a ratio
$\frac{v_1'}{v_1}\equiv x(\tau)+iy(\tau)$, and for convenience we
convent a Wronskian ${\cal W}\equiv v_1'{v_{1}^*}-v_1{v_{1}^*}'$.
Then at the initial conformal time $\tau_0$ the energy density of
the left-handed mode is given by
\begin{eqnarray}
\rho_L=\int_0^\infty\frac{dk k^2{\cal
W}}{16i\pi^2a^4y_0}(x_0^2+y_0^2-2\frac{a'}{a}x_0+\frac{a'^2}{a^2}+k^2)~,
\end{eqnarray}
where $x_0=x(\tau_0)$ and $y_0=y(\tau_0)$ which are the ratio
parameters at the initial conformal time.

Commonly, it is sufficient to obtain the initial conditions by
calculating the variation of the energy density with respect to
$x_0$ and $y_0$. From the vanishing variations we obtain the
``initial" conditions $ x_0=\frac{a'}{a}$ and $y_0^2=k^2$.
However, if these two conditions are satisfied in Eq. (\ref{v1})
and (\ref{v2}), we have to require $\alpha_3$ to be zero which
violates the noncommutativity. Therefore, the lowest energy
density of noncommutative primordial tensor perturbations is not
an extremal point where the variations with respect to $x_0$ and
$y_0$ are zero. Instead, the lowest energy density can only be
chosen on the boundary value of the range we care about. Moreover,
this boundary value should be positive and independent of any
time-dependent phase. Therefore, we deduce that only if $|y_0|$
approaches to $k$ as closely as possible, it is satisfied that the
energy state is the lowest in the range we allow. Consequently, we
obtain the correct initial condition $y_0=-l-4\alpha_3$ for the
left-handed in our note; similarly, we also obtain the
corresponding initial condition for the right-handed.

Thus when $|l\tau|\gg aH$, $v_1$ and $v_2$ are given by
\begin{eqnarray}\label{plane wave}
v_1\rightarrow\frac{1}{\sqrt{2l}}e^{-i(l+4\alpha_3)\tau}~,~~v_2\rightarrow\frac{1}{\sqrt{2l}}e^{-i(l-4\alpha_3)\tau}~,
\label{initial}\end{eqnarray} where $l$ denote the effective
co-moving wave numbers of the modified tensor perturbations, and
its form is given by
\begin{eqnarray}
\label{effective wave
number}l&=&(k^2+16\alpha_3^2)^{\frac{1}{2}}~.
\end{eqnarray}
Thus we see that both $v_1$ and $v_2$ display a well oscillating
behaviour inside of horizon, and are identical with the standard
primordial tensor perturbations if we neglect the noncommutative
terms. During inflation, in the slow roll approximation, we have the
relation
\begin{equation}
\frac{a''}{a}=\frac{\nu^2-\frac{1}{4}}{\tau^2}~,
\end{equation} where
$\nu=\frac{\varepsilon-3}{2(\varepsilon-1)}$ and
$\varepsilon\equiv-\frac{\dot H}{H^2}$ is the slow roll parameter.
Thus with the initial conditions (\ref{initial}), the solutions of
Eqs. (\ref{v1}) and (\ref{v2}) can be given as
\begin{eqnarray}\label{exact v1}
&v_1(k,
\tau)=\frac{\sqrt{\pi}}{2}e^{i(\nu+\frac{1}{2})\frac{\pi}{2}}(-\tau)^{\frac{1}{2}}e^{-4i\alpha_3\tau}H_{\nu}^{(1)}(-l\tau)~,\\
&v_2(k,
\tau)=\frac{\sqrt{\pi}}{2}e^{i(\nu+\frac{1}{2})\frac{\pi}{2}}(-\tau)^{\frac{1}{2}}e^{4i\alpha_3\tau}H_{\nu}^{(1)}(-l\tau)~,
\end{eqnarray}
where $H_{\nu}^{(1)}$ is the $\nu$th Hankel function of the first
kind. It is interesting to notice that the tensor perturbations
obtain an effective mass, and thus in the low energy limit the
effective mass would bring some brand-new phenomena. Consequently,
we conclude that the dispersion relation of primordial gravitational
waves has been modified and locally the Lorentz symmetry is broken.
When the modes $v_1$ and $v_2$ are out of horizon ($|l|\ll aH$), we
can obtain their asymptotic forms:
\begin{eqnarray}\label{h outof horizon1}
&v_1\rightarrow
e^{i(\nu-\frac{1}{2})\frac{\pi}{2}}2^{\nu-\frac{3}{2}}e^{-4i\alpha_3\tau}\frac{\Gamma(\nu)}{\Gamma(\frac{3}{2})}\frac{(-l\tau)^{\frac{1}{2}-\nu}}{\sqrt{2l}}~,\\
\label{h outof horizon2} &v_2\rightarrow
e^{i(\nu-\frac{1}{2})\frac{\pi}{2}}2^{\nu-\frac{3}{2}}e^{4i\alpha_3\tau}\frac{\Gamma(\nu)}{\Gamma(\frac{3}{2})}\frac{(-l\tau)^{\frac{1}{2}-\nu}}{\sqrt{2l}}~,
\end{eqnarray}
which are the solutions of primordial gravitational waves after
adding noncommutative terms. These solutions are used in the next
section.

\section{What can we see from GWB today?}

\subsection{Features of Primordial GWB}

In this section we focus on the connections of theory and
observations, and hence we need to know the detailed features of
primordial gravitational waves. Since the noncommutative terms would
modify the dispersion relation of gravitational waves, it is
interesting to see what would be brought about in the propagations
and spectra of primordial tensor perturbations.




For convenience we define
\begin{eqnarray}
u_1(k,\tau)\equiv\frac{v_1(k,\tau)}{a}~,~~u_2(k,\tau)\equiv\frac{v_2(k,\tau)}{a}~.
\end{eqnarray}
Combining the definition of the power spectrum, we have the
expression:
\begin{eqnarray}\label{PT}
P_T(k,\tau)=32\pi
G\frac{k^3}{2\pi^2}(|u_1(k,\tau)|^2+|u_2(k,\tau)|^2)~.
\end{eqnarray}
Moreover, the power spectra of tensor perturbations at different
time can be connected by a parameter called transfer function.
Therefore, we define the transfer function as follows:
\begin{eqnarray}\label{PTtoday}
P_T(k,\tau)=T(k,\tau)P_T(k,\tau_i)~,
\end{eqnarray}
where the index ($_i$) indicates the end of inflation. To
substitute eqs. (\ref{h outof horizon1}) and (\ref{h outof
horizon2}) into the definition of tensor power spectrum in Eq.
(\ref{PT}), we obtain the analytic solution of primordial power
spectrum $P_T(k)$:
\begin{eqnarray}\label{tensor power slow roll}
P_T(k)\equiv P_T(k,\tau_i)&=&64\pi G\frac{2^{2\nu-3}}{4\pi^2a_i^2}\frac{\Gamma^2(\nu)}{\Gamma^2(\frac{3}{2})}[(1-\varepsilon)a_iH_i]^{2\nu-1}\nonumber\\
&\times&k^3l^{-2\nu}(k)~.
\end{eqnarray}

Furthermore, there have been a number of papers to investigate the
mass of today's graviton, and they provide an upper limit around
$6.395\times10^{-32}$eV
theoretically\cite{Gruzinov01,Cutler02,Cooray03,Jones04}. Taking
into account this limit, we easily find that $\alpha_3$ has to be
less than the frequency of $10^{-16}$Hz, whose Compton wavelength is
in equal order of the size of supercluster. Besides, to require that
the slow-roll parameter $\varepsilon$ is close to zero, we then
derive the form of primordial tensor power spectrum,
\begin{eqnarray}\label{pPT}
P_T(k)\simeq\frac{128\pi}{3M_{pl}^4}V_{inf}\frac{k^3}{(k^2+16\alpha_3^2)^\frac{3}{2}}~.
\end{eqnarray}
If the momentum $k$ is much larger than $\alpha_3$, one can see that
this primordial tensor power spectrum will return to the standard
form $\frac{128\pi}{3M_{pl}^4}V_{inf}$ from the equation above.
However, if $k$ is tuned around the value of NP, the amplitude of
the power spectrum starts to decrease.
In this theory $P_T$ would vanish when we tune $k$ to be zero
which is greatly different from the typical knowledge of GWB.

\begin{figure}
\begin{center}
\includegraphics[width=3.7in]{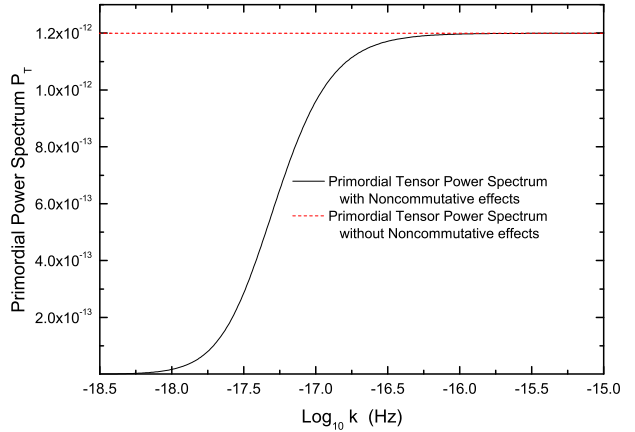}
\end{center}
\caption{The black solid curve represents the primordial power
spectrum of noncommutative tensor perturbations $P_T(k,\tau_i)$ at
the end time of inflation. The red dash curve represents the
primordial power spectrum without noncommutative contributions
while in the same inflation model. In this figure, we adopt the
value of NP as $\alpha_3=10^{-18}$, and the potential of inflation
as $V_{inf}\sim M^4$ in which $M=5\times10^{15}$Gev.}
\label{fig:primordial}
\end{figure}

\begin{figure}
\begin{center}
\includegraphics[width=3.4in]{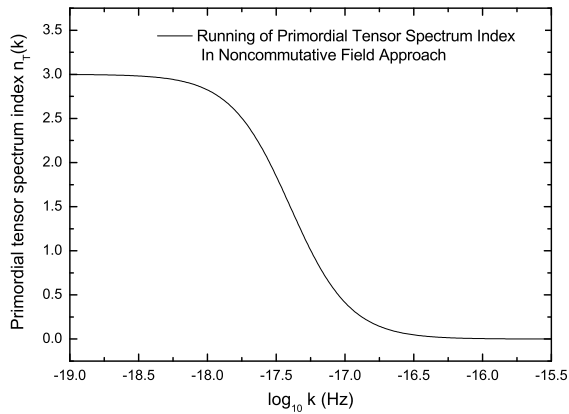}
\end{center}
\caption{The black solid curve depicts the features of primordial
tensor spectral index of noncommutative tensor perturbations
$n_T$. As we know, the tensor spectral index in standard
perturbation theory would reside around zero. However in this
figure it does not stay at zero but runs to 3 which is strongly
different from the standard theory. Here we still adopt
$\alpha_3=10^{-18}$, and the potential of inflation as
$V_{inf}\sim M^4$ in which $M=5\times10^{15}$Gev.} \label{fig:nT}
\end{figure}

Eventually, to make the analysis more specifical, we give a scenario
of primordial tensor power spectrum in the inflation of nearly
constant potential in Fig. \ref{fig:primordial}. Comparing the one
from the standard theory, our primordial tensor power spectrum shows
a greatly dynamical behaviour which is independent of specific
inflation models as shown in Fig. \ref{fig:primordial}. Note that in
Fig. \ref{fig:primordial}, When the value of $k$ is much larger than
NP, the solid line tends forwards the red dash line; when $k$
approaches the NP from the right side, the solid line starts
dropping down and finally vanishes when $k$ approaches zero. The
other way to see this interesting dynamical behaviour of
noncommutative tensor fluctuations is to analyze its spectral index
which is shown in Fig. \ref{fig:nT}. We find that in the range of
frequencies near NP, the spectral index would not reside on the
value $0$ but climb up to a fixed positive value which depends on
the model of inflation.

\subsection{Transfer Function of Gravitational waves}

In the above we discussed the behaviour of noncommutative tensor
perturbations exhibited in primordial tensor power spectrum. Now
what we care about is how to recognize these tensor perturbations in
the GWB nowadays. Since the primordial gravitational waves are
distributed in every frequency, once the effective co-moving wave
number is less than $aH$, the corresponding mode of gravitational
waves would escape the horizon and be frozen until it re-enters the
horizon. The relation between the time when tensor perturbations
leave the horizon and the time  when they return is
$a_{out}H_{out}=a_{in}H_{in}$. Therefore, we have the conclusion
that, the earlier the perturbations escape the horizon, the later
they re-enter it. Moreover, once the effective co-moving wave number
is larger than $aH$, the perturbations begin to oscillate like plane
wave. In the following, we will establish the relation to relate the
power spectrum observed today to the primordial one. The method we
used here is similar to \cite{Steinhardt05}.

In order to make clear every possible ingredient affecting the
evolvement of the GWB, it is suitable and reasonable to decompose
the transfer function into three parts as,
\begin{eqnarray}
T(k,\tau)&=&F_1F_2F_3\nonumber\\
&=&|\frac{\bar
u_{1(2)}(k,\tau)}{u_{1(2)}(k,\tau_i)}|^2|\frac{\tilde
u_{1(2)}(k,\tau)}{\bar
u_{1(2)}(k,\tau)}|^2|\frac{u_{1(2)}(k,\tau)}{\tilde
u_{1(2)}(k,\tau)}|^2.\nonumber\\
\end{eqnarray}
Here $u_{1(2)}(k,\tau)$ is the exact solution of Eq. (\ref{h
equation}); $\tilde u_{1(2)}(k,\tau)$ is an approximate solution of
Eq. (\ref{h equation}) by neglecting the anisotropic stress tensor
$\Sigma_{ij}$; and $\bar u_{1(2)}(k,\tau)$ is an even more crude
solution which is equal to $u_{1(2)}(k,\tau_i)$ if $k<aH$ while
equal to plane wave if $k>aH$. Note that those two polarizations are
only different on their phases which does not affect the expression
of the transfer function. Consequently, we only need to investigate
one polarization in the following.

Firstly, from Eq. (\ref{plane wave}) one can see that after horizon
re-entering, gravitational waves begin to oscillate with a decaying
amplitude proportional to $a^{-1}(\tau)$. Therefore, from the
definition of $\bar u_1$ we get
\begin{eqnarray}
  \bar u_1(k,\tau)= \left\{ \begin{array}{c}
    \frac{u_{1max}(k)}{a(\tau)}\cos[l(\tau-\tau_l)+\phi_l],~~l>aH \\
    \\
    u_1(k,\tau_i),~~l<aH
\end{array} \right.  \label{baru1}
\end{eqnarray}
where $\phi_l$ depends on the initial condition, $u_{1max}(k)$ is
the maximum of the amplitude of oscillation, and $\tau_l$ is the
conformal time when $l=aH$. Since we require this function to be
continuous, there must be a matching relation that
$u_1(k,\tau_i)=[u_{1max}(k)\cos\phi_l]/a(\tau_l)$. Based on these
relations one can get the first factor $F_1$ as follows
\begin{eqnarray}\label{F1L}
F_1=\left(\frac{1+z(\tau)}{1+z_l}\right)^2\cos^2[l(\tau-\tau_l)+\phi_l]/\cos^2\phi_l,
\end{eqnarray}
where we introduce the redshift $1+z=a_0/a(\tau)$ in aim of showing
the effects of the suppression as a result of redshift. The index
``$0$" indicates today, and $z_l$ is the redshift when the modes
re-entered the horizon $l=aH$. Note that the relation of $z_l$ and
$k$ can be given by the following equation
\begin{eqnarray}
\left(\frac{k}{k_0}\right)^2=\sum_i\Omega_i^{(0)}(1+z_1)\exp\left[3\int_0^{z_1}\frac{w_i(\tilde
z)}{1+\tilde z}d\tilde z \right],
\end{eqnarray}
where the sum over $i$ includes all components in the universe.
Since the contributions from dark energy and the fluctuations in
radiation are very small, here we ignore them and then solve out
\begin{eqnarray}\label{zl}
1+z_l=\frac{1+z_{eq}}{2}\left[-1+\sqrt{1+\frac{4(l/k_0)^2}{(1+z_{eq})\Omega_m^{(0)}}}\right],
\end{eqnarray}
where $z_{eq}\equiv-1+\Omega_m^{(0)}/\Omega_r^{(0)}$.
The factor $F_1$ describes the redshift-suppressing effect on the
primordial gravitational waves. Since this factor shows strongly
oscillating behaviour which is inconspicuous to be observed in the
GWB, we usually average the term $\cos^2[l(\tau-\tau_l)+\phi_l]$
and instead it with $\frac{1}{2}$.

Secondly, when considering the influence of the background state
of universe on the re-entry of horizon, we focus on analyzing the
factor $F_2$. Since the background equation of state $w$ varies
very slowly, it is profitable to assume that the evolution of the
scale factor is of form $a=a_0(\frac{\tau}{\tau_0})^{\alpha}$ with
$\alpha=\frac{2}{1+3w}$. Then solving Eq. (\ref{h equation}) again
and ignoring $\Sigma_{ij}$, we have $\tilde
u_1(k,\tau)=u_1(k,\tau_i)\Gamma(\alpha+\frac{1}{2})\left(-\frac{l\tau}{2}\right)^{\frac{1}{2}-\alpha}J_{\alpha-\frac{1}{2}}(-l\tau)$,
where $\Gamma$ is the Gamma function and $J_{\nu}$ is the $\nu$th
Bessel function. If $|l\tau|\gg1$, there is such a relation that
$|\frac{\tilde
u_1(k,\tau)}{u_1(k,\tau_i)}|^2=\frac{\Gamma^2(\alpha+\frac{1}{2})}{\pi}(-\frac{l\tau}{2})^{-2\alpha}\cos^2(l\tau+\frac{\alpha\pi}{2})$.
To match with Eq. (\ref{F1L}), considering that the phase should
be continuous, hence we have the solution that when GWB re-enters
the horizon the conformal time $\tau_l=-\frac{\alpha}{l}$.
Therefore, the second factor $F_2$ is given by
\begin{eqnarray}\label{F2L}
F_2=\frac{\Gamma^2(\alpha+\frac{1}{2})}{\pi}\left(\frac{2}{\alpha}\right)^{2\alpha}\cos^2\phi_l~.
\end{eqnarray}
The second factor shows that, when the gravitational waves re-enter
the horizon, there is a "wall" lying on the horizon which affects
the tensor power spectrum.

Thirdly, during the evolution of tensor perturbations, the nonzero
anisotropic stress tensor $\Sigma_{ij}$ would more or less bring
some effects on the GWB. This effect is proposed by Steven
Weinberg\cite{Weinberg03}, and usually the primary ingredients are
the freely streaming neutrinos which damp the amplitude of the
tensor power spectrum. However, this damping effect just make $P_T$
times a constant but do not change the behaviour of the GWB's
evolving. In our paper, we adopt $F_3=0.80313$.

Ultimately, we have discussed three kinds of leading corrections in
the transfer function which make contributions in the evolution of
the GWB. Using this transfer function, we are able to connect the
primordial gravitational waves with what we observe today.

\subsection{Analysis of Today's GWB}

In this section, we investigate the power spectrum of the current
GWB and the corresponding energy spectrum. To substitute eqs.
(\ref{F1L}), (\ref{F2L}), and the damping factor $F_3$ into
(\ref{PTtoday}), we can give today's tensor power spectrum as
follows,
\begin{eqnarray}\label{tensor power today}
P_T(k,\tau_0)&=&64\pi G\frac{0.80313}{(1+z_l)^2}\frac{\Gamma^2(\alpha+\frac{1}{2})}{2\pi}\left(\frac{2}{\alpha}\right)^{2\alpha}\nonumber\\
&\times&\frac{2^{2\nu-3}}{4\pi^2a_i^2}\frac{\Gamma^2(\nu)}{\Gamma^2(\frac{3}{2})}[(1-\varepsilon)a_iH_i]^{2\nu-1}\frac{k^3}{l^{2\nu}},
\end{eqnarray}
and from Eq. (\ref{approximate tensor energy}), the present energy
spectrum is given by
$\Omega_{GW}(k,\tau_0)=\frac{1}{12}\frac{k^2}{(a_0H_0)^2}P_T(k,\tau_0)$.

\begin{figure}
\begin{center}
\includegraphics[width=3.7in]{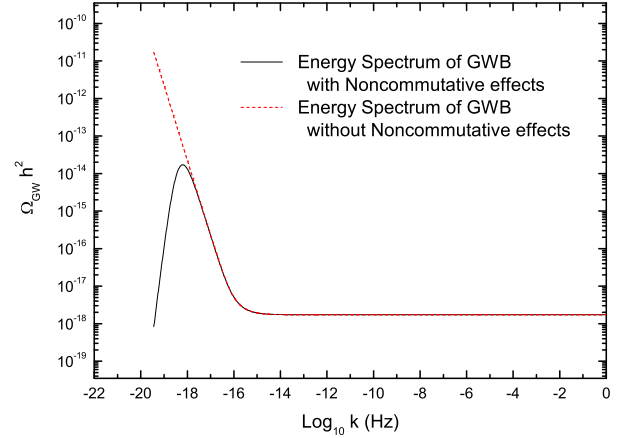}
\end{center}
\caption{The black solid line represents the behaviour of energy
spectrum of noncommutative tensor perturbations $\Omega_{GW}$, and
the red dash line gives the curve of energy spectrum in standard
perturbation theory. The most significant difference is that the
solid line form a peak in low frequency while the red line keeps
raising up. If the NP is small enough, the peak can appear in the
area of CMB Pol's observing. Here we let $\alpha_3=10^{-19}$, and
the potential of inflation be $V_{inf}\sim M^4$ where
$M=5\times10^{15}$Gev.} \label{fig:Omegacom1}
\end{figure}

\begin{figure}
\begin{center}
\includegraphics[width=3.7in]{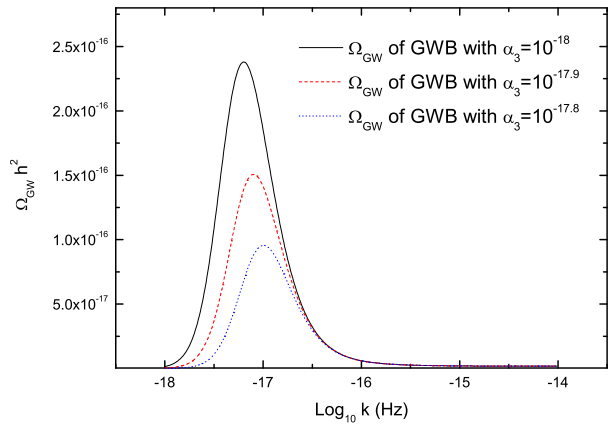}
\end{center}
\caption{The black solid line represents the energy spectrum of
noncommutative tensor perturbations with the smallest NP; the red
dash line gives the curve of energy spectrum with the bigger NP;
and the blue dot curve shows the energy spectrum with the biggest
NP. One can see that the peak of GWB in black solid line is the
most manifest. Here the potential of inflation is taken to be
$V_{inf}\sim M^4$ in which $M=5\times10^{15}$Gev.}
\label{fig:Omegacom3}
\end{figure}

When the frequency of GWB is large enough, one can see that today's
noncommutative tensor perturbation power spectrum would agree with
the standard theory very well. Consequently, there will be very few
signals in the high-frequency range. However, if the frequency
reaches the value near NP where the corresponding wave length is
around the size of the current horizon, the noncommutative term
begins to affect the behaviour of the GWB. Again we require the
slow-roll parameter $\varepsilon$ to tend forwards to zero, and
assume the potential of inflation to be nearly constant of which the
scale is $5\times10^{15}$Gev. Then we give the semi-analytical form
of the present energy spectrum of the noncommutative tensor
perturbations as follows
\begin{eqnarray}\label{Omegasemi}
\Omega_{GW}(k,\tau_0)h^2&=&2ek^5l^{-3}\left(-1+\sqrt{1+fl^2}\right)^{-2}~,
\end{eqnarray}
where $e=2.68563\times10^{14}$ and $f=3.10475\times10^{32}$. In
order to make a comparison, we give the corresponding energy
spectrum without noncommutative term
$\Omega_{GW}^{normal}h^2=2ek^2(-1+\sqrt{1+fk^2})^{-2}$. Note that,
when the frequency of GWB is near NP, the term after the parameter
'$f$' determines the behaviour of the energy spectrum and the term
in the brackets $\left(...\right)$ of Eq. (\ref{Omegasemi}) will
never vanish even $k$ approaches zero. Due to that, the energy
spectrum of noncommutative GWB form a peak in low frequency and
then decay rapidly as mentioned in the beginning. To be more
specifically, we show this feature in Fig. \ref{fig:Omegacom1} and
see that if tuning NP felicitously the peak can be in the
detecting range of next generation of CMB experiments(see CMB Pol
\cite{TFCR}). In Fig. \ref{fig:Omegacom3} we choose three groups
of NP to see the differences among them. We find that, smaller the
NP is, more manifest the peak is. That is to say, it is most
possible to detect the features of noncommutativity in the GWB
with very minor values of NP. This is consistent with the case
that when $\alpha_3$ approaches $0$, the commutative one is
recovered. In fact we select $\alpha_3=10^{-19}$ in Fig.
\ref{fig:Omegacom1} for the same consideration.

\section{Conclusions and Discussions}

As a conclusion, we have investigated the key features of the
primordial tensor perturbations in the noncommutative field
approach, and discussed the possibility of detecting the
corresponding  GWB in the experiments. Due to the noncommutative
effects, the dispersion relation for the primordial gravitational
waves is modified and the solution of the the tensor perturbations
is different from the commutative case. Therefore, it brings about a
lot of exciting phenomena which is brand-new and valuable for us to
investigate. To study these new features, we investigated the
transfer function to obtain the spectrum of noncommutative
gravitational waves that we are observing today. Since the
noncommutative term would bring CPT violation and produce effective
mass for the graviton, it is reasonable to require this term to be
small enough. In our note, this has already been discussed that it
is allowed to set the values of NP lower than $10^{-16}$ due to the
requirements of both experiments and theories. From the calculations
in this note, one can see that one most intriguing effect of
noncommutative GWB is that it would generate a peak on its energy
spectrum where the frequency may be lower than $10^{-16}$. As a
result, on one hand, this phenomenon provides a much more stronger
limit on the graviton mass, since we can check the position of this
possible peak in the energy spectrum; on the other hand, we expect
that the signals of noncommutativity can be found in the next
generation of CMB observations if the noncommutativity in the relic
tensor perturbations is hidden in the range near current horizon.
Eventually, the noncommutativity of gravitational waves definitely
go beyond the knowledge of Einstein's gravity and therefore should
be an important subject for us to investigate.

\acknowledgments

It is a pleasure to thank Xinmin Zhang and Bin Chen for helpful
discussions, and we thank Yi Wang, Gong-Bo Zhao, and Xiao-Fei
Zhang for part of computer calculations. This work is supported in
part by National Natural Science Foundation of China under Grant
Nos. 90303004, 10533010, 19925523 and 10405029, as well as in part
by the Scientific Research Fund of GUCAS(NO.055101BM03).


\vfill

\begin{thebibliography}{99}

\bibitem{Guth81} A. H. Guth, Phys. Rev. {\bf D23}, 347 (1981).

\bibitem{Steinhardt82} A. Albrecht, and P. Steinhardt, Phys. Rev.
Lett. {\bf 48}, 1220 (1982).

\bibitem{Linde82} A. D. Linde, Phys. Lett. {\bf B108}, 389 (1982).

\bibitem{CMBobserve} A. D. Miller {\it et al.}, Astrophys. J. {\bf
524}, L1 (1999), astro-ph/9906421; P. de Bernardis {\it et al.},
Nature {\bf 404}, 955 (2000), astro-ph/0004404; S. Hanany {\it et
al.}, Astrophys. J. {\bf 524}, L5 (2000), astro-ph/0005123; N. W.
Halverson {\it et al.}, Astrophys. J. {\bf 568}, 38 (2002),
astro-ph/0104489; B. S. Mason {\it et al.}, Astrophys. J. {\bf
591}, 540 (2003), astro-ph/0205384; A. Benoit {\it et al.}, Astro.
Astrophys. {\bf 399}, L25 (2003), astro-ph/0210306; J. H.
Goldstein {\it et al.}, Astrophys. J. {\bf 599}, 773 (2003),
astro-ph/0212517;

\bibitem{WMAP} D. N. Spergel {\it et al.}, Astrophys. J. Suppl. {\bf 148}, 175
(2003), astro-ph/0302209; D. N. Spergel {\it et al.},
astro-ph/0603449.

\bibitem{Mukhanov81} V. Mukhanov, and G. Chibisov, JETP {\bf 33}, 549 (1981).

\bibitem{Guth82} A. H. Guth, and S.-Y. Pi, Phys. Rev. Lett. {\bf 49}, 1110
(1982).

\bibitem{Hawking82} S. W. Hawking, Phys. Lett. {\bf B115}, 295 (1982).

\bibitem{Starobinsky82} A. A. Starobinsky, Phys. Lett. {\bf B117}, 175 (1982).

\bibitem{Bardeen83} J. M. Bardeen, P. J. Steinhardt, and M. S. Turner,
Phys. Rev. {\bf D28}, 679 (1983).

\bibitem{BBO} URL: http://universe.nasa.gov/program/vision/bbo.html.


\bibitem{CMBnextobserve} URL: http://www.planck.fr/.

\bibitem{VPJ} L. Verde, H. Peiris, and R. Jimenez,
JCAP {\bf 0601}, 019 (2006), astro-ph/0506036.

\bibitem{Boyle06} L. Boyle, P. Steinhardt, and N. Turok, Phys. Rev. Lett. {\bf 96}, 111301
(2006), astro-ph/0507455.

\bibitem{Smith06} T. Smith, M. Kamionkowski, and A. Cooray, Phys. Rev. {\bf D73}, 023504
(2006), astro-ph/0506422.

\bibitem{Grishchuk75} L. P. Grishchuk, Sov. Phys. JETP {\bf 40},
409 (1975).

\bibitem{Allen88} B. Allen, Phys. Rev. {\bf D37}, 2078 (1988).

\bibitem{Starobinski79} A. Starobinsky, JETP Lett. {\bf 30}, 682
(1979); V. Rubakov, M. Sazhin, and A. Veryaskin, Phys. Lett. {\bf
B115}, 189 (1982); R. Fabbri, and M. Pollock, Phys. Lett. {\bf
B125}, 445 (1983); L. Abbott, and M. Wise, Nucl. Phys. {\bf B244},
541 (1984); L. Abbott, and D. Harari, Nucl. Phys. {\bf B264}, 487
(1986).

\bibitem{Lyth93} E. Stewart, and D. Lyth, Phys. Lett. {\bf B302}, 171
(1993), gr-qc/9302019.

\bibitem{Gasperini93} M. Gasperini, and M. Giovannini, Phys. Rev. {\bf
D47}, 1519 (1993), gr-qc/9211021; R. Brustein, M. Gasperini, M.
Giovannini, and G. Veneziano, Phys. Lett. {\bf B361}, 45 (1995),
hep-th/9507017; A. Buonanno, M. Maggiore, and C. Ungarelli, Phys.
Rev. {\bf D55}, 3330 (1997), gr-qc/9605072.

\bibitem{Boyle04} L. Boyle, P. Steinhardt, and N. Turok, Phys. Rev.
{\bf D69}, 127302 (2004), hep-th/0307170.

\bibitem{piao06} Y.-S. Piao, Phys. Rev. {\bf D73}, 047302 (2006),
gr-qc/0601115.

\bibitem{piao03} Y.-S. Piao, and E. Zhou, Phys. Rev. {\bf D68}, 083515
(2003), hep-th/0308080; Y.-S. Piao, and Y.-Z. Zhang, Phys. Rev.
{\bf D70}, 063513 (2004), astro-ph/0401231.

\bibitem{pin-others} P. F. Gonzalez-Diaz, and J. A. Jimenez-Madrid, Phys. Lett. {\bf B596}, 16 (2004),
hep-th/0406261; M. Baldi, F. Finelli, and S. Matarrese, Phys. Rev.
{\bf D72}, 083504 (2005), astro-ph/0505552.


\bibitem{Douglas97} A. Connes, M. Douglas, and A. Schwarz, JHEP {\bf
9802}, 003 (1998), hep-th/9711162.

\bibitem{Seiberg99} N. Seiberg, and E. Witten, JHEP {\bf 9909}, 032 (1999),
hep-th/9908142.

\bibitem{Bigatti00} D. Bigatti, and L. Susskind, Phys. Rev.
{\bf D62}, 066004 (2000), hep-th/9908056; N. Seiberg, L. Susskind,
and N. Toumbas, JHEP {\bf 0006}, 021 (2000), hep-th/0005040.

\bibitem{Alekseev00} A. Yu. Alekseev, A. Recknagel, and V. Schomerus, JHEP {\bf
0005}, 010 (2000), hep-th/0003187.

\bibitem{Chu02} C.-S. Chu, and P.-M. Ho, Nucl. Phys. {\bf B636}, 141 (2002),
hep-th/0203186.

\bibitem{Seiberg00} N. Seiberg, L. Susskind, and N. Toumbas, JHEP
{\bf 0006}, 044 (2000), hep-th/0005015.

\bibitem{Gomis00} J. Gomis, and T. Mehen, Nucl. Phys. {\bf B591}, 265 (2000),
hep-th/0005129.

\bibitem{Carroll01} S. M. Carroll, J. Harvey, V. Kostelecky, C. Lane, and T.
Okamoto, Phys. Rev. Lett. {\bf 87}, 141601 (2001), hep-th/0105082.

\bibitem{Carson02} C. Carlson, C. Carone, and R. Lebed, Phys. Lett. {\bf B549},
337 (2002), hep-ph/0209077.

\bibitem{Chaichian04} M. Chaichian, P. Kulish, K. Nishijima, and A. Tureanu, Phys.
Lett. {\bf B604}, 98 (2004), hep-th/0408069.



\bibitem{Brandenberger00}  R. Brandenberger, and P.-M. Ho, Phys. Rev. {\bf D66}, 023517
(2002), hep-th/0203119; S. Tsujikawa, R. Maartens, and R.
Brandenberger, Phys. Lett. {\bf B574}, 141 (2003),
astro-ph/0308169; G. Calcagni, Phys. Rev. {\bf D70}, 103525
(2004), hep-th/0406006.

\bibitem{HuangLi} Q. Huang, and M. Li, JHEP {\bf 0306}, 014 (2003),
hep-th/0304203; Q. Huang, and M. Li, JCAP {\bf 0311}, 001 (2003),
astro-ph/0308458; Q. Huang, and M. Li, Nucl. Phys. {\bf B713}, 219
(2005), astro-ph/0311378.

\bibitem{Cai04} R.-G. Cai, Phys. Lett. {\bf B593}, 1 (2004), hep-th/0403134.

\bibitem{HuangZhang} Q.-G. Huang, Phys. Rev. {\bf D74} 063513 (2006), astro-ph/0605442;
X. Zhang, JCAP {\bf 0612}, 002 (2006), hep-th/0608207.

\bibitem{Fatollahi} A. H. Fatollahi, and M. Hajirahimi, Phys. Lett. {\bf B641}, 381
(2006), hep-th/0611225.

\bibitem{BY} K. Bamba, and J. Yokoyama, Phys. Rev.
{\bf D70}, 083508 (2004), hep-ph/0409237.


\bibitem{Carmona02} J. M. Carmona, J. L. Cortes, J. Gamboa, and F.
Mendez, Phys. Lett. {\bf B565}, 222 (2003), hep-th/0207158.

\bibitem{Gamboa05} J. Gamboa, and J. Lopez-Sarrion, Phys. Rev. {\bf D71}, 067702 (2005), hep-th/0501034; H.
Falomir, J. Gamboa, J. Lopez-Sarrion, F. Mendez, and A. J. da
Silva, Phys. Lett. {\bf B632}, 740 (2006), hep-th/0504032.

\bibitem{Nascimento06} J. R. Nascimento, A. Petrov, and R. F. Ribeiro,
hep-th/0601077.

\bibitem{Arias06} P. Arias, A. Das, J. Gamboa, J. Lopez-Sarrion, and F. Mendez,
hep-th/0610152.

\bibitem{Ferrari06} A. F. Ferrari, M. Gomes, J. R. Nascimento, E. Passos, A. Yu.
Petrov, and A. J. da Silva, hep-th/0609222.

\bibitem{Feng06} B. Feng, M. Li, J. Xia, X. Chen, and X. M. Zhang,
Phys. Rev. Lett. {\bf 96}, 221302 (2006), astro-ph/0601095.


\bibitem{Weinberg03} S. Weinberg, Phys. Rev. {\bf D69}, 023503
(2004), astro-ph/0306304.

\bibitem{Pritchard04}  J. R. Pritchard, and M. Kamionkowski, Annals Phys. {\bf 318}, 2
(2005), astro-ph/0412581.

\bibitem{Bashinsky} S. Bachinsky, astro-ph/0505502.

\bibitem{Dicus05} D. A. Dicus, and W. Repko, Phys . Rev. {\bf D72}, 088302
(2005), astro-ph/0509096.

\bibitem{Brandenberger92} V. F. Mukhanov, H. A. Feldman, and R. H.
Brandenberger, Phys. Rept. {\bf 215}, 203 (1992).

\bibitem{BrandenbergerM00} J. Martin, and R. H. Brandenberger, Phys. Rev. {\bf D63}, 123501
(2001), hep-th/0005209; J. Martin, and R. Brandenberger, Phys.
Rev. {\bf D68}, 063513 (2003), hep-th/0305161.

\bibitem{Gruzinov01} A. Gruzinov, New Astron. {\bf 10}, 311 (2005),
astro-ph/0112246.

\bibitem{Cutler02} C. Cutler, W. Hiscock, and S. Larson, Phys. Rev. {\bf D67},
024015 (2003), gr-qc/0209101.

\bibitem{Cooray03} A. Cooray, and N. Seto, Phys. Rev. {\bf D69}, 103502 (2004),
astro-ph/0311054.

\bibitem{Jones04} D. I. Jones, Astrophys. J. {\bf 618}, L115 (2004),
gr-qc/0411123.

\bibitem{Steinhardt05} L. A. Boyle, and P. J. Steinhardt, astro-ph/0512014.

\bibitem{TFCR} R. Weiss {\it et al.}, Task Force On Cosmic Microwave
Research, (www.science.doe.gov/hep/TFCRreport.pdf).


\end{thebibliography}
\end{document}